\documentclass[conference,a4paper]{IEEEtran}

\usepackage{cite}
\usepackage{graphicx,color,epsfig,rotating}
\usepackage{amsfonts,amsmath,amssymb,bbm}
\usepackage{algorithm}
\usepackage{algpseudocode}
\usepackage{subfigure}
\usepackage{amsmath}
\usepackage{cite}
\usepackage{mdwtab}
\usepackage{subfigure}
\usepackage{placeins}
\usepackage{psfrag, graphicx}
\usepackage[latin1]{inputenc}
\usepackage{amssymb}
\usepackage{makeidx}
\usepackage{epstopdf}
\usepackage{enumitem}

\long\def\comment#1{}

\newfont{\bbb}{msbm10 scaled 700}

\newfont{\bb}{msbm10 scaled 1100}

\newcommand{\be}{\begin{equation}}
\newcommand{\ee}{\end{equation}}
\newcommand{\bea}{\begin{eqnarray}}
\newcommand{\eea}{\end{eqnarray}}

\newtheorem{theorem}{Theorem}
\newtheorem{claim}{Claim}
\newtheorem{remark}{Remark}

\begin{document}

\title{Heterogeneous Computation Assignments in Coded Elastic Computing}

\author{Nicholas Woolsey, Rong-Rong Chen and Mingyue Ji
\thanks{The authors are with the Department of Electrical Engineering,
University of Utah, Salt Lake City, UT 84112, USA. (e-mail: nicholas.woolsey@utah.edu, rchen@ece.utah.edu and mingyue.ji@utah.edu)}
}

\author{
    \IEEEauthorblockN{ Nicholas Woolsey,
		Rong-Rong Chen, and Mingyue Ji }
	\IEEEauthorblockA{Department of Electrical and Computer Engineering, University of Utah\\
		Salt Lake City, UT, USA\\
		Email: \{nicholas.woolsey@utah.edu,
		 rchen@ece.utah.edu,
		mingyue.ji@utah.edu\}}

}

\maketitle

\thispagestyle{empty}
\pagestyle{empty}

\begin{abstract}
We study the optimal design of a heterogeneous coded elastic computing (CEC) network where machines have varying relative computation speeds. CEC introduced by Yang {\it et al.} is a framework which mitigates the impact of elastic events, where machines join and leave the network. A set of data is distributed among storage constrained machines using a Maximum Distance Separable (MDS) code such that any subset of machines of a specific size can perform the desired computations. This design eliminates the need to re-distribute the data after each elastic event. In this work, we develop a process for an arbitrary heterogeneous computing network to minimize the overall computation time by defining an optimal computation load, or number of computations assigned to each machine. We then present an algorithm to define a specific computation assignment among the machines that makes use of the MDS code and meets the optimal computation load.
\end{abstract}


\section{Introduction}
\label{section: intro}

Coding has been proposed as an effective tool to speed up computations of distributed computing networks. Examples include Coded Distributed Computing (CDC) for MapReduce-like distributed computing platforms \cite{li2018fundamental}  and coded data shuffling used in distributed machine learning applications \cite{attia2019shuffling,elmahdy2018shuffling,wan2018fundamental}, where codes are designed to significantly  minimize the communication load by increasing the computation capability and/or the storage size on each machine. Another example is to use codes to mitigate the straggler effect in applications such as matrix multiplications \cite{lee2017speeding,yu2018straggler}, where any subset of machines with a cardinality larger than the recovery threshold can recover the matrix multiplication. This eliminates the need to wait for the computation of slow machines.

Similar to straggler mitigation coded computing designs, Coded Elastic Computing (CEC) was introduced by Yang {\it et al.} in 2019 to mitigate {\it preempted} machines \cite{yang2018coded}. In this framework, a storage limited computing network performs computations over many time steps. Between each time step an elastic event may occur where machines become preempted (unavailable) or become available again. Computations are performed on a set of data, for example a matrix, and the computations change each time step. For example, in each time step the data matrix may be multiplied with a different vector. In each time step, the goal becomes to assign computations among the available machines. A naive approach is to assign each machine a non-overlapping part of the data. However, this is inefficient as the storage has to be redefined with each elastic event. 

The idea of CEC is to use a Maximum Distance Separable (MDS) code to distribute coded data among the machines. The data is split into $L$ equal sized, disjoint data sets and each machine stores a coded combination of these sets. In this way, each machine only stores an equivalent of an $\frac{1}{L}$ fraction of the data. Furthermore, any computation can be resolved by combining the coded computation results of $L$ machines. Then, given a set of available machines the coded computations are assigned to the machines such that each computation is assigned to $L$ machines. In the original CEC scheme of \cite{yang2018coded}, the authors proposed a ``cyclic" computation assignment such that each machine is assigned the same number computations.

The recent work \cite{dau2019optimizing} also studies CEC and aims to maximize the overlap of the task assignments between computation time steps. With each elastic event, the computation assignment must change. In the cyclic approach in \cite{yang2018coded}, the assignments in the current time step are independent of assignments in previous time steps. In \cite{dau2019optimizing}, the authors design assignment schemes to minimize the changes in the assignments between time steps. In some cases, the proposed assignment schemes were shown to achieve zero transition waste, or minimize the amount of new local computations at the machines. However, both \cite{yang2018coded} and \cite{dau2019optimizing} only study homogeneous computing networks.


In this paper, we propose a CEC framework optimized for a heterogenous network where machines have varying computation speeds. In this setting, more computations are assigned to faster machines and less computations to slower machines to minimize the maximum local computation time among the machines. This assignment problem is non-trivial since by the MDS code design we still require that each computation is assigned to $L$ machines. We propose and solve an optimization problem to find the optimal computation load, or amount of computations assigned to each machine. We then show an assignment exists that yields this computation load and design a low complexity algorithm
to find such an assignment.\footnote{ The CEC assignment algorithm is adapted from our heterogeneous private information retrieval (PIR) storage placement algorithm of \cite{woolsey2019optimal}.} Our proposed CEC design works for an arbitrary set of machine speeds and requires a number of computation assignments at most equal to the number of available machines.

\paragraph*{Notation Convention}
We use $|\cdot|$ to represent the cardinality of a set or the length of a vector 
and $[n] := [1,2,\ldots,n]$. 

\section{Network Model and Problem Formulation}
\label{sec: Network Model and Problem Formulation}

We consider a set of $N$ machines. Each stores a coded matrix derived from a $q\times r$ data matrix, $\boldsymbol{X}$. The coded matrices are defined by an $N\times L$ MDS generator matrix $\boldsymbol{G}=(g_{n,\ell})$ 
such that any $L$ rows of $\boldsymbol{G}$ are invertible. The data matrix, $\boldsymbol{X}$, is row-wise split into $L$ disjoint, $\frac{q}{L}\times r$ matrices, $\boldsymbol{X}_1,\ldots ,\boldsymbol{X}_L$. Each  machine $n\in[N]$ stores the $\frac{q}{L}\times r$ coded matrix
\be
\boldsymbol{\tilde{X}}_n = \sum_{\ell=1}^{L}g_{n,\ell}\boldsymbol{X}_\ell.
\ee

The machines collectively perform matrix-vector computations over multiple times steps. In a given time step only a subset of the $N$ machines are available to perform matrix computations. More specifically, in time step $t$, a set of available machines $\mathcal{N}_t \subseteq [N]$ aims to compute
\be
\boldsymbol{y}_t = \boldsymbol{X}\boldsymbol{w}_t
\ee
where $\boldsymbol{w}_t$ is some vector of length $r$. The machines of $[N]~\setminus~\mathcal{N}_t$ are preempted and we assume the number of available machines $N_t=|\mathcal{N}_t| \geq L$ as at least $L$ machines are assumed to be available in each time step.

The machines of $\mathcal{N}_t$ do not compute $\boldsymbol{y}_t$ directly. Instead, each machine $n\in \mathcal{N}_t$ computes the set
\be
\mathcal{V}_{n} = \left\{ v = \boldsymbol{\tilde{X}}_n^{(i)}\boldsymbol{w}_t : i \in \mathcal{W}_{n} \right\}
\ee
where $\boldsymbol{\tilde{X}}_n^{(i)}$ is the $i$-th row of $\boldsymbol{\tilde{X}}_n$ and $\mathcal{W}_{n}\subseteq\left[ \frac{q}{L}\right]$ is the set of rows assigned to machine $n$ in time step $t$. Furthermore, we define the computation load vector, $\boldsymbol{\mu}$, such that
\be \label{eq: compload_vector}
\mu[n] = \frac{|\mathcal{W}_{n}|}{
\left( \frac{q}{L} \right)}, \;\; \forall n \in \mathcal{N}_t
\ee
is the fraction of rows computed by machine $n$ in time step $t$. Note that, $\boldsymbol{\mu}$, $\mathcal{V}_{n}$ and $\mathcal{W}_{n}$ change with each time step, but reference to $t$ is omitted for ease of disposition. Moreover, the machines have varying computation speeds defined by the strictly positive vector, $\boldsymbol{s}$, which is fixed over all time steps. Here, computation speed is the number of row multiplications per unit time. The computation time is dictated by the machine that takes the most time to perform its assigned computations such that the computation time in a particular time step is
\be
c(\boldsymbol{\mu}) = \max_{n\in \mathcal{N}_t} \frac{\mu[n]}{s[n]}.
\ee

In a given time step, for each $i\in\left[ \frac{q}{L}\right]$, $L$ machines perform the vector-vector multiplication with the $i$-th row of their local coded matrix and $\boldsymbol{w}_t$. The results are sent to a master node which can resolve the elements of $\boldsymbol{y}_t$ by the MDS code design. To assign each row to $L$ machines, we define $F$ disjoint sets of rows, $\boldsymbol{\mathcal{M}}_t  = (\mathcal{M}_{1},\ldots,\mathcal{M}_{F} )$ whose union is $\left[ \frac{q}{L}\right]$. Then, $F$ sets of $L$ machines, $\boldsymbol{\mathcal{P}}_t = (\mathcal{P}_{1} , \ldots , \mathcal{P}_{F} )$, are defined such that $\mathcal{P}_f \subseteq\mathcal{N}_t$ and $|\mathcal{P}_f|=L$ for all $f\in [F]$. The rows of $\mathcal{M}_{f}$ are assigned to the machines of $\mathcal{P}_{f}$. The rows computed by machine $n\in \mathcal{N}_t$ in time step $t$ are in the set
\be
\mathcal{W}_{n} = \bigcup \left\{ \mathcal{M}_f : f\in[F],  n \in \mathcal{P}_{f}\right\}
\ee
and $\boldsymbol{\mu}$ is a function of $\left(\boldsymbol{\mathcal{M}}_t, \boldsymbol{\mathcal{N}}_t \right)$. The sets $\mathcal{M}_{1},\ldots,\mathcal{M}_{F}$ and $\mathcal{P}_{1} , \ldots , \mathcal{P}_{F}$ and $F$ may vary with each time step.

In a given time step $t$, our goal is to define the computation assignments, $\boldsymbol{\mathcal{M}}_t$ and $\boldsymbol{\mathcal{P}}_t $, such that the resulting computation load vector defined in (\ref{eq: compload_vector}) has the minimum computation time.
In time step $t$, given $\mathcal{N}_t$ and $\boldsymbol{s}$, the optimal computation time, $c^*$, is the infimum of computation time defined by all possible computation assignments, $\left(\boldsymbol{\mathcal{M}}_t, \boldsymbol{\mathcal{P}}_t \right)$, such that
\begin{align}
c^* = &\inf_{\left(\boldsymbol{\mathcal{M}}_t, \boldsymbol{\mathcal{P}}_t \right)}c\left(\boldsymbol{\mu}\left(\boldsymbol{\mathcal{M}}_t, \boldsymbol{\mathcal{N}}_t \right)\right) \nonumber\\
&\text{s.t. }\bigcup_{\mathcal{M}_f \in \boldsymbol{\mathcal{M}}_t}\mathcal{M}_f=\left[ \frac{q}{L}\right],\label{eq: optprob_assign}\\
&\;\;\;\;\;\; |\mathcal{P}_f|= L \;\; \forall \mathcal{P}_f \in \boldsymbol{\mathcal{P}}_t,\nonumber\\
&\;\;\;\;\;\; |\boldsymbol{\mathcal{M}}_t|=|\boldsymbol{\mathcal{P}}_t|. \nonumber
\end{align}

It can be seen that the optimization problem (\ref{eq: optprob_assign}) is combinatorial such that the optimal solution and the optimal value are non-trivial. In Sections~\ref{sec: compload} and \ref{sec: compassign}, we solve this combinatorial optimization problem by decomposing it into two sub-problems: 1) a convex optimization problem to find an optimal $\boldsymbol{\mu}$ without the consideration of a specific computation assignment and 2) a computation assignment problem. Moreover, we show that an optimal assignment, $\left(\boldsymbol{\mathcal{M}}_t, \boldsymbol{\mathcal{P}}_t \right)$, can be found via a low complexity algorithm.

\begin{figure*}
\centering
\centering \includegraphics[width=18.25cm]{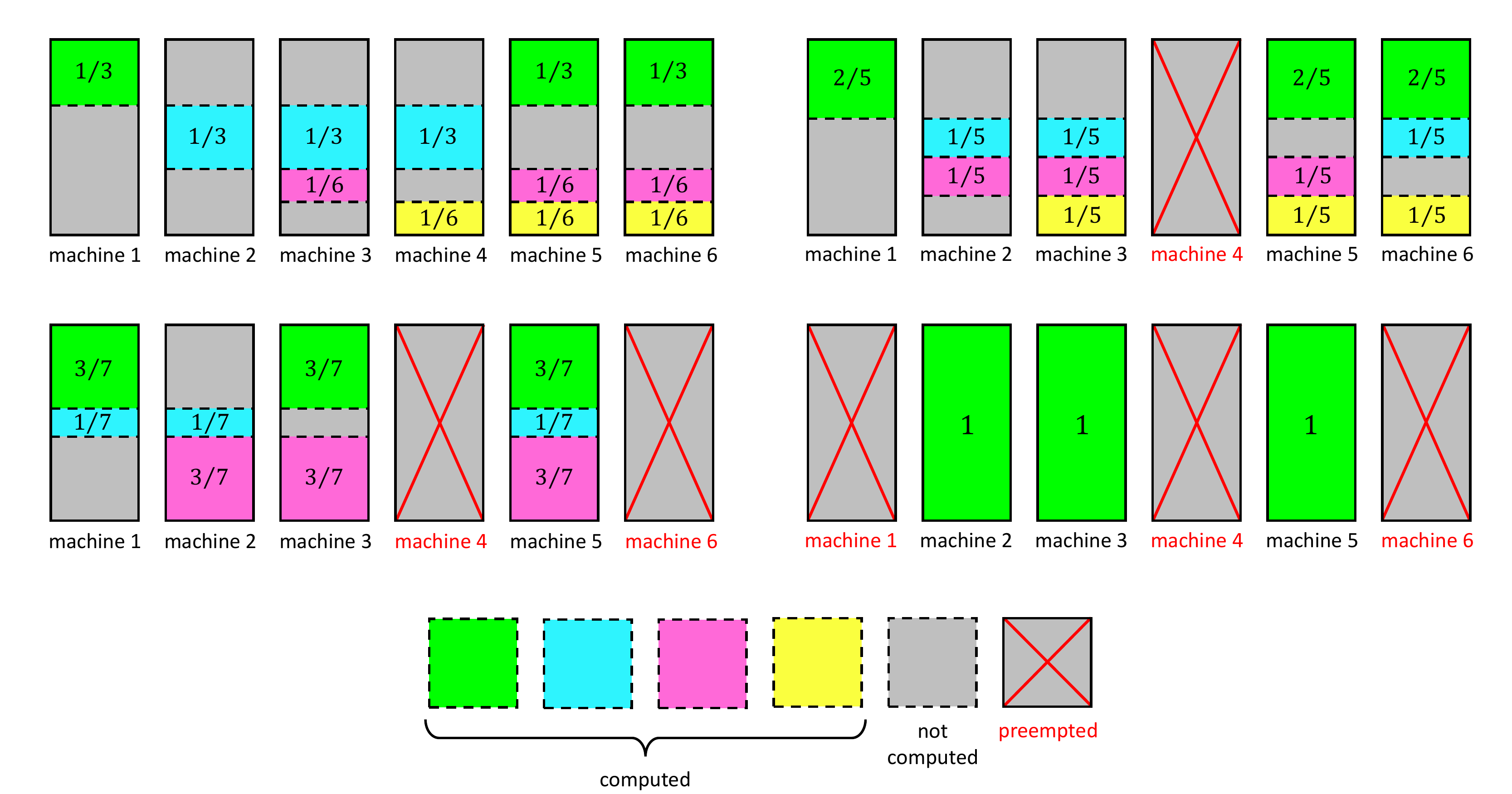} 
\put(-155,59.5){a) no preempted machines, $t=1$}
\put(-63.5,59.5){b) one preempted machine, $t=2$}
\put(-155,25){c) two preempted machines, $t=3$}
\put(-63.5,25){d) three preempted machines, $t=4$}
\put(-98,-3){e) legend}
\put(-128.5,14.5){{\large$\mathcal{M}_1$}}
\put(-114.75,14.5){{\large$\mathcal{M}_2$}}
\put(-101,14.5){{\large$\mathcal{M}_3$}}
\put(-87.25,14.5){{\large$\mathcal{M}_4$}}
\vspace{-0.cm}
\caption{~\small Optimal computation assignments over $4$ times steps on a heterogeneous CEC network.}
\label{fig: exp1}
\vspace{-0.4cm}
\end{figure*}
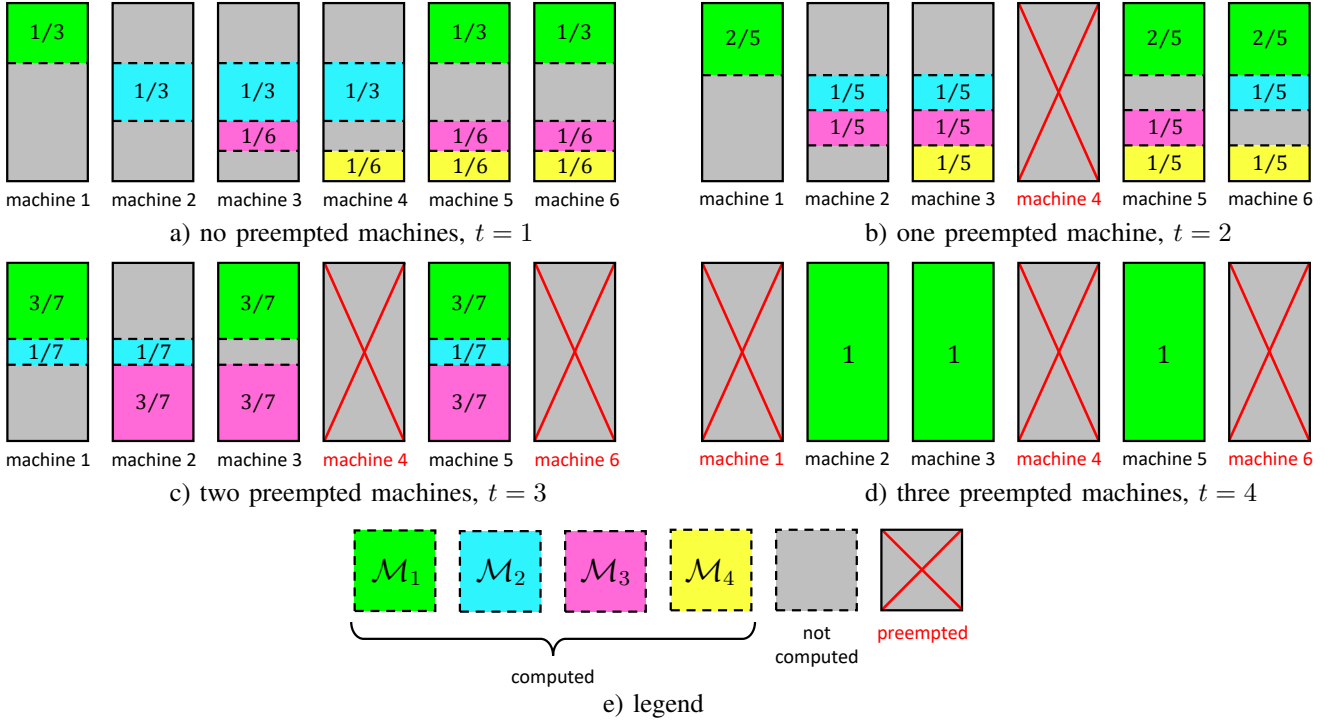

\section{An Example}

\label{sec: example}

There are a total of $N=6$ machines where each has the storage capacity to store $\frac{1}{3}$ of a data matrix $\boldsymbol{X}$. In time step $t$, the machines have the collective goal of computing $\boldsymbol{y}_t=\boldsymbol{X}\boldsymbol{w}_t$ where $\boldsymbol{w}_t$ is some vector. In order to allow for preempted machines, $\boldsymbol{X}$ is split row-wise into $L=3$ sub-matrices, $\boldsymbol{X}_1$, $\boldsymbol{X}_2$, and $\boldsymbol{X}_3$ and a MDS code is used to define the matrices $\{\boldsymbol{\tilde{X}}_n : n\in[N] \}$ which are stored among the machines. 
This placement is designed such that any element of $\boldsymbol{y}_t$ can be recovered by obtaining the corresponding coded computation from any $3$ machines. For example, the first element of $\boldsymbol{y}_t$ can be recovered from the results of machines $1$, $3$ and $5$ multiplying the top row of their respective coded matrix with $\boldsymbol{w}_t$. To recover the entirety of $\boldsymbol{y}_t$, we split the coded matrices into sets of rows, such that each set is used for computation at $L=3$ machines.

The machines have relative computation speeds defined by
\be
\boldsymbol{s} = [\;2,\;\;2,\;\;3,\;\;3,\;\;4,\;\;4\;].
\ee
Machines $5$ and $6$ are the fastest machines and can perform row computations twice as fast as machines $1$ and $2$. Machines $3$ and $4$ are the next fastest machines and can perform matrix computations $1.5$ times as fast as machines $1$ and $2$. Our goal is to assign computations, or rows of the coded matrices, to the machines to minimize the overall computation time such that each computation is assigned to $3$ machines.

In time step $1$, there are no preempted machines and $\mathcal{N}_1~=~\{1,\ldots, 6 \}$. We assign fractions of the rows to the machines defined by the computation load vector
\be
\boldsymbol{\mu} = \left[\;\frac{1}{3},\;\;\frac{1}{3},\;\;\frac{1}{2},\;\;\frac{1}{2},\;\;\frac{2}{3},\;\;\frac{2}{3}\;\right]
\ee
such that machines $1$ and $2$ are assigned $\frac{1}{3}$, machines $3$ and $4$ are assigned $\frac{1}{2}$ and machines $5$ and $6$ are assigned $\frac{2}{3}$ of the rows of their respective coded matrices. We define $\boldsymbol{\mu}$ such that it sums to $L=3$ and each row can be assigned to $3$ machines. Furthermore, based on the machine computation speeds, the machines finish at the same time to minimize the overall computation time. In Section \ref{sec: compload}, we will discuss 
the systematic approach to determine $\boldsymbol{\mu}$. Next, given $\boldsymbol{\mu}$, the rows of the coded matrices must be assigned. We define sets of rows, $\mathcal{M}_{1}$, $\mathcal{M}_{2}$, $\mathcal{M}_{3}$, and $\mathcal{M}_{4}$ which are assigned to sets of machines $\mathcal{P}_{1}$, $\mathcal{P}_{2}$, $\mathcal{P}_{3}$, and $\mathcal{P}_{4}$, respectively. These sets are depicted in Fig.~\ref{fig: exp1}(a) where $\mathcal{M}_{1}$ contains the first $\frac{1}{3}$ of the rows which are assigned to machines $\mathcal{P}_{1} = \{ 1,5,6\}$ and $\mathcal{M}_{2}$ contains the next $\frac{1}{3}$ of the rows and is assigned to machines $\mathcal{P}_{2}=\{2,3,4\}$. Moreover, $\mathcal{M}_{3}$ contains the next $\frac{1}{6}$ of the rows are assigned to machines $\mathcal{P}_{3}=\{3,5,6\}$ and $\mathcal{M}_{4}$ contains the final $\frac{1}{6}$ of the rows are assigned to machines $\mathcal{P}_{4}=\{4,5,6\}$. Later in Section \ref{sec: compassign}, we present Algorithm \ref{algorithm:1}, which defines the computation assignment for general $\boldsymbol{\mu}$. By this assignment, the fraction of rows assigned to machine $i$ sums to $\mu[i]$ and each row is assigned to $L=3$ machines so that the entirety of $\boldsymbol{y}_1$ is recovered.

In time step $2$, $N_t=5$ as machine $4$ is preempted and is no longer available to perform computations. Therefore, the computations must be reassigned. First, we define
\be
\boldsymbol{\mu} = \left[\;\frac{2}{5},\;\;\frac{2}{5},\;\;\frac{3}{5},\;\;0,\;\;\frac{4}{5},\;\;\frac{4}{5}\;\right]
\ee
which sums to $L=3$ and minimizes the overall computation time. Given $\boldsymbol{\mu}$, we then use Algorithm \ref{algorithm:1}, which aims to assign computations to a machine with the least remaining rows to be assigned and $L-1=2$ machines with the most remaining rows to be assigned. For example, in the first iteration, $\mathcal{M}_{1}$ is defined to contain the first $\frac{2}{5}$ rows and is assigned to machines $\mathcal{P}_{1}= \{ 1,5,6\}$. After this iteration, machines $3$, $5$ and $6$ require $\frac{2}{5}$ of the total rows to still be assigned to them and machine $3$ requires $\frac{3}{5}$ of the total rows. In the next iteration, $\mathcal{M}_{2}$ contains the next $\frac{1}{5}$ of the rows and is assigned to $\mathcal{P}_{2}=\{ 2,3,6\}$. Note that, only $\frac{1}{5}$ of the rows could be assigned in this iteration otherwise there would only be two machines, $3$ and $5$, which still require assignments and therefore, the remaining rows cannot be assigned to three machines. In the final two iterations, $\mathcal{M}_{3}$ and $\mathcal{M}_{4}$ contain $\frac{1}{5}$ of the previously unassigned rows and are assigned to the machines of $\mathcal{P}_{3} = \{2,3,5 \}$ and $\mathcal{P}_{4} = \{3,5,6 \}$, respectively. These assignments are depicted in Fig.~\ref{fig: exp1}(b).

Next, in time step $3$, machines $4$ and $6$ are preempted. Similar to previous examples it is ideal to have machines $3$ and $5$ compute $1.5\times$ and $2\times$ the number of computations, respectively, compared to machines $1$ and $2$. However, this is not possible since each machine can be assigned at most a number of rows equal to the number of rows of the coded matrices. In this case, we assign all rows to the fastest machine, machine $5$, and assign fractions of the rows to the remaining machines which sum up to $2$. As a result, we define
\be
\boldsymbol{\mu} = \left[\;\frac{4}{7},\;\;\frac{4}{7},\;\;\frac{6}{7},\;\;0,\;\;1,\;\;0\;\right].
\ee
Then, Algorithm \ref{algorithm:1} defines, $\mathcal{M}_{1}$, $\mathcal{M}_{2}$ and $\mathcal{M}_{3}$, disjoint sets containing $\frac{3}{7}$, $\frac{1}{7}$ and $\frac{3}{7}$ of the rows respectively. Moreover, these row sets are assigned to the machines of $\mathcal{P}_{1}= \{ 1,3,5\}$, $\mathcal{P}_{2}= \{ 1,2,5\}$ and $\mathcal{P}_{3}= \{ 2,3,5\}$, respectively. These assignments are depicted in Fig.~\ref{fig: exp1}(c).

Finally, in time step $4$, machines $1$, $4$ and $6$ are preempted. To assign all the rows to $L=3$ machines, each available machine is assigned all of the rows and
\be
\boldsymbol{\mu} = \left[\;0,\;\;1,\;\;1,\;\;0,\;\;1,\;\;0\;\right].
\ee
In other words, $\mathcal{M}_{1}$ contains all rows and $\mathcal{P}_{1}=\{2,3,5 \}$. This is depicted in Fig.~\ref{fig: exp1}(d).

\section{Optimal Computation Load Vector}
\label{sec: compload}
In this section, we  introduce a relaxed optimization problem of (\ref{eq: optprob_assign}) that is convex and solve it to find the optimal computation load vector $\boldsymbol{\mu^*}$ directly from the speed vector $\bold s$  without considering the computation assignment $(\boldsymbol{\mathcal{M}}_t,\boldsymbol{\mathcal{P}}_t)$ explicitly. In Section \ref{sec: compassign}, we will show that there exists a computation assignment $(\boldsymbol{\mathcal{M}}_t,\boldsymbol{\mathcal{P}}_t)$ that yields the optimal computation load vector $\boldsymbol{\mu^*}$ found by the relaxed optimization problem.
Throughout the remainder of this paper, without loss of generality, we assume that $\mathcal{N}_t = \{1,2,\ldots , N_t \}$ where $N_t $ is the number of available machines in time step $t$. We ignore the computation load of any preempted or unavailable machine which is simply $0$.

\subsection{A Relaxed Convex Optimization Problem}

Given a computation speed vector $\bold s$, we define the optimal computation load vector $\boldsymbol{\mu}^*$ to be the solution to the following relaxed optimization problem:
\begin{align}
\boldsymbol{\mu}^* = \;&\underset{\boldsymbol{\mu}}{\mathrm{argmin}}\max_{n\in [N_t]}\frac{\mu[n]}{s[n]} \nonumber\\
&\;\text{s.t.} \sum_{n\in [N_t]}\mu[n] = L \label{eq: optprob}\\
&\;\;\;\;\; 0 \leq \mu[n] \leq 1, \forall n \in [N_t] \nonumber,
\end{align}
which can be shown to be convex.
While computation assignments, $(\boldsymbol{\mathcal{M}}_t,\boldsymbol{\mathcal{P}}_t)$, are not explicitly considered in (\ref{eq: optprob}), we note that the key constraint of $\sum_{n\in [N_t]}\mu[n] = L$ is a relaxed version of that requirement on the computation assignment  that each row should be assigned to $L$ machines.
When $N_t=L$, the solution to (\ref{eq: optprob}) is $\boldsymbol{\mu}^* = [1,\dots, 1]$. The analytical solution to (\ref{eq: optprob}) when $N_t>L$ is presented in Theorem \ref{th: load_assignment}.

\begin{theorem}\label{th: load_assignment}
Assume that $N_t>L$ and  $s[1] \leq s[2] \leq \cdots \leq s[N_t]$.
    The optimal solution $\boldsymbol \mu^*$ to  the optimization problem of (\ref{eq: optprob}) must take the following form
\begin{equation}
\mu^*[n]=\begin{cases}
\hat c^* s[n] & \text{if } 1\le n \le k^*\\
1 & \text{if } k^*+1 \le n \le N_t,
\end{cases}
\label{eq:optimal_form}
\end{equation}
where $k^*$ is the largest integer in $[N_t-L+1, N_t]$ such that
\be
\frac{1}{s[k^*+1]} < \hat c^* = \frac{k^*+L-N_t}{\sum_{n=1}^{k^*}s[n]} \leq \frac{1}{s[k^*]}.
\label{eq: c_bounds-fork}
\ee
Here, ${\hat c}{^*}=c(\boldsymbol \mu^*)$ is the maximum computation time among the $N_t$ machines given the computation load assignment  $\boldsymbol \mu^*$.
The left side of (\ref{eq: c_bounds-fork}) is ignored when $k^*=N_t$.
\end{theorem}

\begin{IEEEproof}

\begin{claim}\label{cl: 1}
If $\boldsymbol \mu^*$ is an optimal solution to (\ref{eq: optprob}), then for every $n\in[N_t]$ we must have either  $ \mu^*[n] =\hat{c}^*  s[n] $ or $1\!=\!\mu^*[n]<\hat{c}^* s[n]$,  where ${\hat c}{^*}=c(\boldsymbol \mu^*)$.
\end{claim}

We prove Claim \ref{cl: 1} by contradiction.
Since $\hat c^*=\text{max}_{n \in N_t}\frac{\mu^*[n]}{s[n]}$, we define two disjoint sets  $\mathcal{T}_0 \bigcup \mathcal{T}_1=[N_t]$ such that \be
\mathcal{T}_0=\{n\in[N_t]:\mu^*[n]=\hat{c}^*  s[n]\}
\ee
and
\be
\mathcal{T}_1=\{n\in[N_t]:\mu^*[n]<\hat{c}^*  s[n]\}.
\ee
Assume for there exists some $i\in[N_t]$ such that $ i \in \mathcal{T}_1$ and $\mu^*[i] < 1$.
Define $\boldsymbol{\mu}'$ such that
\begin{align}
\mu'[n] = \left\{ \begin{array}{cc}
                \mu^*[n] + \epsilon  \hspace{5mm} &\text{if }n=i, \\
                \mu^*[n] - \frac{\epsilon}{|\mathcal{T}_0|}  \hspace{5mm} &\text{if }  n\in\mathcal{T}_0, \\
                \mu^*[n]  \hspace{5mm} &\text{if } n \in  \mathcal{T}_1 \setminus i\\
                \end{array} \right.
\end{align}
where $\epsilon >0$ is sufficiently small such that
\be
\frac{\mu'[i]}{s[i]} = \frac{\mu^*[i]+\epsilon}{s[i]} < \hat{c}^*
\ee
and for all $n \in \mathcal{T}_0$
\be
 \mu^*[n] - \frac{\epsilon}{|\mathcal{T}_0|} > 0.
\ee
One can verify that we have $\frac{\mu'[n]}{s[n]} < \hat{c}^*$ for any $n \in [N_t]$ and thus we obtain $c(\mu') < \hat{c}^*$. This contradicts with the assumption that $\boldsymbol{\mu}^*$ is optimal. Thus, it follows that if $n \notin \mathcal{T}_0$, then we must have $n\in \mathcal{T}_1$ and $\mu^*[n]=1$.

\begin{claim}\label{cl: 2}
  If $j \in \mathcal{T}_0 $ and $i \in \mathcal{T}_1$, then
   $s[j] < s[i]$.
\end{claim}
This follows from
\be\label{eq: pfeq1}
\frac{\mu^*[i]}{s[i]}=\frac{1}{s[i]} < \hat{c}^*= \frac{\mu^*[j]}{s[j]} \leq \frac{1}{s[j]}.
\ee

Combining Claims \ref{cl: 1} and \ref{cl: 2}, we find that the optimal solution must take the form of
\begin{equation}
\mu^*[n]=\begin{cases}
\hat c_k^* s[n] & \text{if } 1\le n \le k\\
1 & \text{if } k+1 \le n \le N_t,
\end{cases}
\label{eq:optimal_form_k}
\end{equation}
where $k=|\mathcal{T}_0|$. Next, we will optimize $k$ such that $\hat c_k^*$ is minimized.
Combining (\ref{eq:optimal_form_k}) and (\ref{eq:optimal_form_k}) we obtain (\ref{eq: c_bounds-fork}) since
\begin{align}
L = \sum_{n=1}^{N_t}\mu^*[n]
  &= N_t - k + \sum_{n=1}^{k}\mu^*[n] \\
  &= N_t - k + \hat{c}_k^*\sum_{n=1}^{k}s[n].
\end{align}
The left-most inequality of (\ref{eq: c_bounds-fork}) follows from  $k \in \mathcal{T}_0$ and $\mu^*[k] \le 1$. The right-most inequality  of (\ref{eq: c_bounds-fork}) follows from  $k+1 \in \mathcal{T}_1$ and $\mu^*[k+1] = 1$.
 Since $\bold s$ is an increasing sequence, we see from  (\ref{eq: c_bounds-fork}) that $\hat c_k^*$ is maximized when $k$ is chosen to be $k^*$,  the largest value in $[N_t-L+1, N_t]$ such that (\ref{eq: c_bounds-fork}) is satisfied.
\end{IEEEproof}

\begin{remark}
The two cases in (\ref{eq:optimal_form}) are determined by whether a machine $n$ satisfies $ \mu^*[n]=\hat c^* s[n]$ or $ \mu^*[n]< \hat c^* s[n]$. For $1 \le n \le k^*$, the equality is achieved and we must have $0< \mu^*[n]\le 1$. When $k^*+1 \le n \le N$, we have the strict  inequality and $ \mu^*[n]=1$.
 The equality in (\ref{eq: c_bounds-fork}) ensures that $\sum_{n=1}^{N_t} \mu^*[n]=L$; the right-most inequality ensures that $\mu^*[n] \le \mu^*[k^*]=\hat c^* s[k^*] \le 1,$ for any $1\le n \le k^*$;  the left-most inequality ensures that for any $k^*+1 \le n \le N$, we have
 $ \mu^*[n]< \hat c^* s[n]$. Hence, the worst computation time $\hat c^*$ is induced by the  $k^*$ slowest machines.
\end{remark}

Since the optimization problem of (\ref{eq: optprob}) aims to minimize a convex function on a closed and convex set, the existence of an optimal solution is guaranteed. This ensures the existence of
  some $k \in [N_t-L+1, N_t]$ such that (\ref{eq: c_bounds-fork}) is satisfied. In the following, we provide a numerical procedure to find $k^*$.  First, it is straightforward to verify that if the right-hand-side (RHS) inequality ``$\le $'' of (\ref{eq: c_bounds-fork}) is violated for $k=i$, then the  left-hand-side (LHS) inequality ``$<$'' of (\ref{eq: c_bounds-fork}) must hold for $k=i-1$. In other words, for any $i=N_t, \!\cdots, \!N_t\!-\!L+2$,
\be
\text{If } \hat c_i^* > \frac{1}{s[i]}, \text{then } \frac{1}{s[i]} < \hat c_{i-1}^*.
\label{eq:equivalent}
\ee
To demonstrate the existence of such a $k$, we first check $k=N_t$. If the RHS of (\ref{eq: c_bounds-fork}) holds, then we have $k^*=N_t$. Otherwise, it follows from (\ref{eq:equivalent}) that the LHS of (\ref{eq: c_bounds-fork}) must hold for $k=N_t-1$. If the RHS of (\ref{eq: c_bounds-fork}) also hold for $k=N_t-1$, then we have $k^*=N_t-1$. Otherwise, it follows from (\ref{eq:equivalent}) that the LHS of (\ref{eq: c_bounds-fork}) must hold for $k=N_t-2$. We continue this process by decreasing $k$ until we find one value of $k$ for which both sides of (\ref{eq: c_bounds-fork}) hold. This process is guaranteed to terminate before reaching $k=N_t-L+1$ for which the RHS of (\ref{eq: c_bounds-fork}) always hold. Hence, this establishes the procedure to find $k^*$ directly using (\ref{eq: c_bounds-fork}).

\subsection{Computation Load Examples}

We return to the first example and explain how to find the optimal computation load vector. When $t=1$, we have $N_t=6, L=3$.
Given $\boldsymbol{s} = [2,\;2,\;3,\;3,\;4,\;4]$, one can verify that the largest $k$ that satisfies (\ref{eq: c_bounds-fork}) is $k^*=6$, and thus $\hat c^*=1/6$,  $\boldsymbol{\mu^*}= \hat c^* \boldsymbol{s}=\left[\frac{1}{3},\frac{1}{3},\frac{1}{2},\frac{1}{2},\frac{2}{3},\frac{2}{3}\right]$.
 Similarly, for $t=2$, since machine 4 preempts, we have now $N_t=5$, and $\boldsymbol{s} = [2,\;2,\;3,\;4,\;4]$ (we ignore any preempted machines). In this case, we have $k^*=5$, and thus $\hat c^*=1/5$,  $\boldsymbol{\mu^*}= \hat c^* \boldsymbol{s}=\left[\frac{2}{5},\frac{2}{5},\frac{3}{5}, \frac{4}{5},\frac{4}{5}\right].$
Similarly, for $t=3$, we have $N_t=4$, and $\boldsymbol{s} = [2,\;2,\;3,\;4]$ because machines 4 and 6 preempts.
Here, we have  $k^*=3$, $\hat c^*=2/7$, and $\boldsymbol{\mu^*}=\left[\frac{4}{7},\frac{4}{7},\frac{6}{7}, 1\right].$ 
Note that, similar to the optimization problem of (\ref{eq: optprob}), the computation load of the preempted machines are ignored since they are simply $0$, presenting a slight difference between the optimal computation load vectors presented in Section \ref{sec: example}.
\section{Optimal Computation assignment}
\label{sec: compassign}

In this section, we show that a computation assignment, $(\boldsymbol{\mathcal{M}}_t,\boldsymbol{\mathcal{P}}_t)$, exists that yields the computation load vector, $\boldsymbol{\mu}^*$, and therefore is an optimal assignment. Moreover, we will provide an iterative algorithm to define such an assignment.

Our goal is to assign computations among the machines such that each computation is assigned to $L$ machines and the assignments satisfy $\boldsymbol{\mu}^*$. This is equivalent to the filling problem (FP) introduced in \cite{woolsey2019optimal} and necessary and sufficient conditions were derived for the existence of the solution.
In particular, a solution exists {\it if and only if}
\begin{align}
\mu^*[n] \leq \frac{\sum_{i=1}^{N_t}\mu^*[i]}{L}
\end{align}
for all $n \in [N_t]$. In this case, we see that $\sum_{i=1}^{N_t}\mu^*[i]=L$ and $\mu^*[n]\leq 1$ for all $n \in [N_t]$. Therefore, an optimal computation assignment exists. Moreover, we provide Algorithm \ref{algorithm:1} to define the optimal computation assignment, $\left(\boldsymbol{\mathcal{M}}_t, \boldsymbol{\mathcal{P}}_t \right)$.\footnote{Algorithm \ref{algorithm:1} is adapted from our previous work \cite{woolsey2019optimal} for storage placement in private information retrieval. More details of Algorithm \ref{algorithm:1} are found in \cite{woolsey2019optimal}.}

\begin{algorithm}
  \caption{Computation Assignment: Heterogeneous CEC}
  \label{algorithm:1}
  \begin{algorithmic}[1]
  \item[ {\bf Input}: $\boldsymbol{\mu}^*$, $N_t$, $L$, and $q$ ]
  \item $\boldsymbol{m} \leftarrow \boldsymbol{\mu}^*$
  \item $f \leftarrow 0$
  \While {$\boldsymbol{m} > \boldsymbol{0}$}
    \State $f \leftarrow f+1$
    \State $L' \leftarrow \sum_{n=1}^{N_t}m[n]$
    \State $\boldsymbol{\ell} \leftarrow$ indices of non-zero elements of $\boldsymbol{m}$ from smallest to largest
    \State $N'\leftarrow$ number of non-zero elements in $\boldsymbol{m}$
    \State $\mathcal{P}_{f} \leftarrow\{\ell [1], \ell [N'-L+2] , \ldots , \ell [N'] \}$
    \If {$N' \geq L+1$}
    \State $\alpha_f \leftarrow  \min \left(\frac{L'}{L} - m[\ell[N' - L + 1]], m[\ell[1]]\right)$
    \Else
    \State $\alpha_f \leftarrow  m[\ell[1]]$
    \EndIf
    \For {$n \in \mathcal{P}_{f}$}
    \State $m[n] \leftarrow m[n] - \alpha_f$
    \EndFor
  \EndWhile
  \item $F \leftarrow f$
  \State Partition rows $[\frac{q}{L}]$ into $F$ disjoint row sets: $\mathcal{M}_{1}, \ldots , \mathcal{M}_{F}$ of size $\frac{\alpha_1 q}{L},\ldots,\frac{\alpha_{F}q}{L}$ rows respectively
  \end{algorithmic}
\end{algorithm}

\begin{remark}
In \cite{woolsey2019optimal}, Algorithm \ref{algorithm:1} was shown to take at most $N_t$ iterations to complete. Therefore, $F \leq N_t$ and at most there are $N_t$ computations assignments.
\end{remark}

\subsection{Example Using Algorithm \ref{algorithm:1}}

\begin{figure}
\centering
\centering \includegraphics[width=8cm]{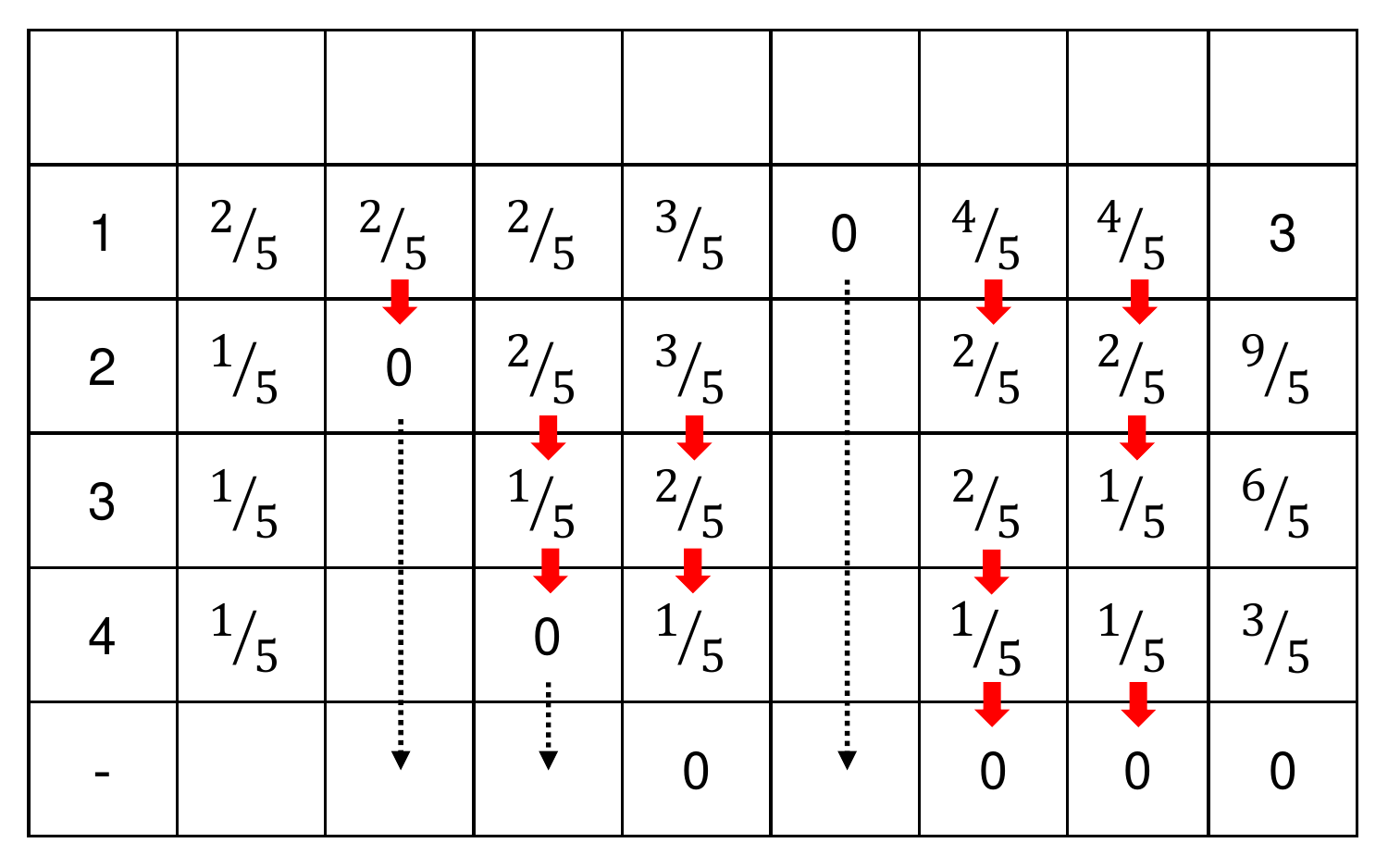} 
\put(-75,43.2){$f$}
\put(-67.5,43.2){$\alpha_f$}
\put(-60,43.2){$m[1]$}
\put(-51.5,43.2){$m[2]$}
\put(-43,43.2){$m[3]$}
\put(-34.5,43.2){$m[4]$}
\put(-26,43.2){$m[5]$}
\put(-17.5,43.2){$m[6]$}
\put(-8,43.2){$L'$}
\vspace{-0.2cm}
\caption{~\small Task assignment by Algorithm \ref{algorithm:1} for example of \ref{sec: example} $(t=2)$.}
\label{table: example}
\vspace{-0.4cm}
 \end{figure}

We return to the example of Section \ref{sec: example} and use Algorithm \ref{algorithm:1} to derive the computation (row) assignments for $t=2$. The steps of the algorithm are shown in Fig.~\ref{table: example}. In the first iteration, $f=1$, $\boldsymbol{m}=\boldsymbol{\mu}$ as no computations have been assigned yet. Rows of the respective coded matrices are assigned to machine $1$, which is a machine with the least remaining computations to be assigned, and machines $5$ and $6$ with the most remaining computations to be assigned. Moreover,
\be
m[1] = \frac{2}{5} \leq \frac{L'}{L} - m[3] = 1 - \frac{3}{5}=\frac{2}{5}
\ee
where machine $3$ is the machine with the most remaining rows to be assigned that is not included in $\mathcal{P}_1=\{1,5,6\}$. Therefore, a fraction $\alpha_1=\frac{2}{5}$ of the rows are assigned to machines ${1,5,6}$. Then, $\boldsymbol{m}$ is adjusted to reflect the remaining computations to be assigned and $L'=3-3\alpha_1 = \frac{9}{5}$.

In the second iteration, $f=2$, machine $2$ is a machine with the least remaining rows to be assigned. Computations are assigned to machine $2$ and machines $3$ and $6$ which are a pair of machines with the most remaining computations to be assigned. Ideally, we would like to assign all the remaining rows to machine $2$. However,
\be
m[2] = \frac{2}{5} > \frac{L'}{L} - m[5] = \frac{3}{5} - \frac{2}{5}=\frac{1}{5}
\ee
and assigning the remaining rows to machine $2$ in this iteration will prevent a valid solution going forward. Therefore, $\alpha_2=\frac{1}{5}$ and after this iteration $\boldsymbol{m}$ and $L'$ are adjusted accordingly.

In the third iteration, $f=3$,
\be
m[2] = \frac{1}{5} \leq \frac{L'}{L} - m[6] = \frac{2}{5} - \frac{1}{5} = \frac{1}{5}
\ee
and an $\alpha_3=\frac{1}{5}$ of the rows are assigned to machines ${2,3,5}$. $\boldsymbol{m}$ and $L'$ are adjusted accordingly. Finally, in the fourth iteration, $f=4$, the three machines with remaining assignments, machines ${3,5,6}$ are assigned an $\alpha_4=\frac{1}{5}$ of the rows. After the $4$ iterations, $m[n]=0$ for all $n\in{1,2,3,4}$ and the computation assignment is complete.

\section{Conclusion}
\label{sec: conclusion}
In this work, we study coded elastic computing where machines store MDS coded data and have varying computation speed. Given a set of available machines with arbitrary relative computation speeds, we derive an optimal computation load among the machines. Then, we show the existence of a computation assignment which yields the optimal computation load. The assignment makes use of the MDS code design by assigning computations to $L\in\mathbb{Z}^+$ machines. Moreover, we present a low complexity algorithm to define the computation assignments with at most a number of iterations equal to the number of available machines. Our coded elastic computing design has the potential to perform computations faster than the state-of-the-art design which was developed for a homogeneous computing network.

\appendices

\bibliographystyle{IEEEbib}
\bibliography{references_d2d}

\end{document}